\documentclass[preprint,twocolumn]{aastex701}

\usepackage{longtable,bm,amsmath}
\usepackage{CJKutf8, color}
\usepackage{soul,multirow}
\usepackage{tikz}

\renewcommand{\figurename}{Fig.}

\shorttitle{Future Detection of NS 1987A
}
\shortauthors{Dohi et al.}
\begin{document}

\title{NS 1987A: A Potential Probe of Crustal Superfluidity with Future X-ray Observations}

\author[0000-0001-8726-5762]{Akira Dohi}
\email[]{dohi@riken.jp}
\affiliation{Astrophysical Big-Bang Laboratory (ABBL), Pioneering Research Institute (PRI), RIKEN, Wako, Saitama 351-0198, Japan}
\affiliation{RIKEN Center for Interdisciplinary Theoretical and Mathematical Sciences (iTHEMS), 2-1 Hirosawa, Wako, Saitama 351-0198, Japan}

\author{Keita Izumi}\email[]{}
%\email[]{2013131s@gsuite.kobe-u.ac.jp}
\affiliation{Department of Physics, Kobe University, Kobe 657-8501, Japan}

\author{Asuka Ito}\email[]{}
%\email[]{asuka@phys.sci.kobe-u.ac.jp}
\affiliation{Department of Physics, Kobe University, Kobe 657-8501, Japan}

\author{Emanuele Greco}\email[]{}
\affiliation{Anton Pannekoek Institute for Astronomy, University of Amsterdam, Science Park 904, 1098 XH Amsterdam, The Netherlands}
\affiliation{GRAPPA, University of Amsterdam, Science Park 904, 1098 XH Amsterdam, The Netherlands}
\affiliation{INAF-Osservatorio Astronomico di Palermo, Piazza del Parlamento 1, 90134 Palermo, Italy}
\affiliation{Universit\`a degli Studi di Palermo, Dipartimento di Fisica e Chimica, Piazza del Parlamento 1, 90134 Palermo, Italy}

\author{Marco Miceli}\email[]{}
\affiliation{Universit\`a degli Studi di Palermo, Dipartimento di Fisica e Chimica, Piazza del Parlamento 1, 90134 Palermo, Italy}
\affiliation{INAF-Osservatorio Astronomico di Palermo, Piazza del Parlamento 1, 90134 Palermo, Italy}

\author{Salvatore Orlando}\email[]{}
\affiliation{INAF-Osservatorio Astronomico di Palermo, Piazza del Parlamento 1, 90134 Palermo, Italy}

\author{Shigehiro Nagataki}\email[]{}
%\email[]{shigehiro.nagataki@riken.jp}
\affiliation{Astrophysical Big Bang Laboratory (ABBL), RIKEN Pioneering Research Institute (PRI), 2-1 Hirosawa, Wako, Saitama 351-0198, Japan}
\affiliation{RIKEN Center for Interdisciplinary Theoretical and Mathematical Sciences (iTHEMS), 2-1 Hirosawa, Wako, Saitama 351-0198, Japan}
\affiliation{RIKEN-Berkeley Center, RIKEN Center for Interdisciplinary Theoretical and Mathematical Sciences (iTHEMS), 2-1 Hirosawa, Wako, Saitama 351-0198, Japan}
\begin{abstract} 

Several indirect observations imply that the central compact object in the supernova SN 1987A is likely to be the youngest neutron star identified to date, NS 1987A. We investigate how a future detection of its thermal X-ray emission could constrain neutron-star interiors, using cooling models and a generic high-resolution, large-area X-ray observatory with \textit{Lynx}-like capabilities as a sensitivity benchmark. For the adopted sensitivity and observing window, NS 1987A is detectable at $t\sim60~{\rm yr}$ (roughly 2047) unless rapid neutrino cooling operates, such as the direct Urca process. If the neutron superfluidity in the inner crust is weak, the temperature profile inside NS 1987A is likely not to be isothermal at the detection. In that case, the future direct observation of NS 1987A is a potential probe of the properties inside the inner crust. We compare alternative X-ray observatories and emphasize that both effective area and angular resolution, needed to reduce contamination from the surrounding remnant, determine whether this diagnostic can be realized.

%This implies that the future direct observation of NS 1987A is a potential probe to the properties of the core in case of strong neutron superfluidity, while the inner crust in case of weak superfluidity.

%the inner crust in weak 

%strong crust superfluidity, while the inner crust is in weak.

%This implies that the future direct observation of NS 1987A is a potential probe to the neutron-star core, but not the crust, although 

%Thermal relation timescale tends to be shorter as the neutron superfluidity in the inner crust is weaker

\end{abstract}


\section{Introduction} \label{sec:intro}

%%Introduction of SN 1987A and NS 1987A
The central compact object (CCO) in supernova (SN) 1987A, which was confirmed through the detection of neutrinos by Kamiokande~\citep{1987PhRvL..58.1490H,1987PhRvL..58.1494B}, is a unique astrophysical laboratory because of its youth, enabling us to study the formation of supernova remnant (SNR 1987A) and a neutron star (NS 1987A). Thanks to such attractions, various wavelength observations have been operated for more than 30 years. 
%\eg{I would suggest some modification in this paragraph in terms of structure. Instead of listing the explanation of the blob seen by JWST, we could just list all the observed emission features and close the paragraph stating that this even JWST observation cannot discriminate between a thermal or a nonthermal origin. In the following I report my suggested new text:}

Multi-wavelength observations across the electromagnetic spectrum, have detected emission probably associated with NS 1987A. ALMA observations~\citep{2019A&A...624A.116U} have revealed a blob of reprocessed emission possibly stemming from NS 1987A. NuSTAR observations \citep{2021ApJ...908L..45G,gmo22} indicated that the hard ($\gtrsim 8$ keV) X-ray spectra of SN1987A require a nonthermal emission component, likely originating from a very young pulsar wind nebula (PWN). Black-body radiation is emitted by NS 1987A if a thick layer of light elements encloses the NS (e.g. \citealt{2020ApJ...898..125P}). More recently, JWST ~\citep{2024Sci...383..898F} have found an excess of emission associated with either this thermal emission from NS 1987A or with the nonthermal emission from the PWN. All these findings imply that the CCO is likely to be the youngest neutron star. Naturally, it is possible that the emission observed by JWST and ALMA is due to a combination of nonthermal and thermal emission, but for our goals, we will focus on the latter.
%\eg{I would remove the whole next section until the paragraph starting with "In Dohi et al. (2023) ..."}

%In particular, recent observations of ALMA~\citep{2019A&A...624A.116U} and JWST~\citep{2024Sci...383..898F} have found the blob probably associated with NS 1987A, which implies that the CCO is likely to be the youngest neutron star (NS 1987A).

%%Two scenario for NS 1987A

%There are two candidates of physical origin to explain the observed blob in SN 1987A (see Table A in \cite{2020ApJ...898..125P} for details). One is the non-thermal emission scenario, more concretely, synchrotron radiation induced by the spin-down of NS 1987A. This power-law component accounts for the NuSTAR spectrum around 8 keV, in particular in 2012 and 2014~\citep{2021ApJ...908L..45G} (but see also \cite{2021ApJ...916...76A}), though this signature seems not strong in the case of 2020~\citep{gmo22}. The other one is a thermal emission scenario, that is, black-body radiation emitted on the surface of NS 1987A. This scenario holds if the envelope of NS 1987A is composed of a thick layer of light elements \cite{2020ApJ...898..125P}. Our paper focuses on the latter case, i.e., thermal emission scenario.

%\eg{I suggest the remove the previous text until here}

%%Dohi et al. 2023
In \cite{2023ApJ...949...97D} (hereafter Paper I), we showed the diagnostic potential of thermal X-ray observations of NS 1987A using the  \textit{Lynx} concept, with a characteristic angular resolution of 0.5 arcsec and a mirror effective area of $\sim2~{\rm m}^2$ at a photon energy of 1 keV, as a benchmark. \textit{Lynx} currently has no funded implementation or scheduled launch, and may never be realized; it should therefore not be regarded as an expected observing facility. NASA is reassessing the concept for the next decadal survey rather than preparing it for launch~\citep{2026arXiv260900033L}. Hereafter, ``\textit{Lynx}-like'' denotes a generic observatory with the benchmark imaging and sensitivity capabilities adopted in Paper I, without implying a particular mission or launch date. Paper I compared cooling models with limiting intrinsic luminosities at $t=50~{\rm yr}$ (2037), including absorption by supernova ejecta~\citep{oon20}. Extending this comparison to later ages allows us to study how the epoch of a future thermal detection would constrain the stellar interior, independently of whether \textit{Lynx} itself is launched.

%when thermal emission from NS 1987A is directly detected

Through the comparison between the observed NS luminosity or surface temperature and the theoretical one based on so-called cooling theory, one can extract information on neutron-star interior, such as the EOS and nucleon superfluidity (for reviews, see \cite{2004ARA&A..42..169Y,2006NuPhA.777..497P,2021PrPNP.12003879B}). While most observed isolated NSs are older than the relaxation time scale as $t_{\rm relax}\sim10-100~{\rm yrs}$\citep{1994ApJ...425..802L,2001MNRAS.324..725G,2020A&A...642A..42S}, they are likely to be isothermal \citep{2011MNRAS.411.1977Y}. On the other hand, NS 1987A is comparable with $t_{\rm relax}$. Due to its youth, the crust may be hotter than the core, implying that neutrino emission in the inner crust is significant due to its strong temperature dependence. In fact, Paper I showed the importance of crust superfluidity for the detectability by a \textit{Lynx}-like observatory. As the detection age is delayed, the neutrino emission in the core becomes more significant, such as the nucleon direct Urca (DU) process, which is a rapid cooling process strongly dependent on the EOS~\citep{1991PhRvL..66.2701L}. Thus, the future detection of NS 1987A must provide various information on NS properties depending on the detection age.

%the fast neutrino cooling process.

%SN 1987A is a more promising unique candidate to probe the matter inside the inner crust. 

%\cite{2019NatSR...918477T} of recent ultra-cold atom gas studies).

%For example, the observed cooling rate of Cassiopeia A presents the first evidence of neutron superfluidity in the core with $T_c\sim5\times10^8~{\rm K}$, where $T_c$ is the critical temperature \citep{2011PhRvL.106h1101P,2011MNRAS.412L.108S}. A recent paper also considered the possibility that Cassiopeia A causes the DU process under moderate neutron superfluidity~\citep{2022MNRAS.511.5843L,2026JHEAp..4900441P}.

%\ka{Paper Iではdirect Urcaを気にしていなかったことをどこに書こう。。。}
% implying that they are likely to be isothermal \citep{2011MNRAS.411.1977Y}., NS 1987A may not be the case, i.e., the crust may be hotter than the core. If the detection age is earlier 

In this work, we investigate the thermal evolution of NS 1987A and the information on its interior that could be obtained from a future thermal X-ray detection. We use the \textit{Lynx}-like sensitivity benchmark to calculate illustrative detection ages, including models with the DU process, which Paper I did not consider. The physical connection between crustal superfluidity, thermal relaxation, and the observable luminosity is independent of the telescope; the numerical detection ages depend on its sensitivity and availability. We discuss alternative observatories and their limitations in Section~\ref{sec:con}.

\section{Cooling models and observations}
\label{sec:obs}

%%Setup of cooling
To describe thermal luminosity, there are basically four kinds of parameters: EOS, mass $M_{\rm NS}$, nucleon superfluidity, and envelope mass $M_{\rm env}$. 
%In this work, we employ two EOSs of APR~\citep{1998PhRvC..58.1804A}, BSk21~\citep{2010PhRvC..82c5804G,2013A&A...560A..48P}, and Y+K~\citep{2021PhLB..82036587M,2022PTEP.2022i3D03M,2025PhRvC.111d5802M}, which have different radii, 11.6 km, 12.6km, and 13.2 km at $1.4~M_\odot$ stars, respectively.
In this work, we employ two EOSs of APR~\citep{1998PhRvC..58.1804A} and Y+K~\citep{2021PhLB..82036587M,2022PTEP.2022i3D03M,2025PhRvC.111d5802M}, which have different radii, 11.6 km and 12.7 km at $1.40~M_\odot$ stars, respectively.
% The corresponding threshold masses of the DU process are $M_{\rm DU}/M_\odot=$1.96, 1.62 and 1.31, respectively. 
The corresponding threshold masses of the DU process are $M_{\rm DU}/M_\odot=$1.96 and 1.31, respectively. We consider only light NSs with $M_{\rm NS}\simeq1.2$--$1.6~M_\odot$ according to model comparison with observed supernova lightcurves of SN1987A~(see \cite{2020ApJ...898..125P} for details). Hence, the cooling models with APR EOS always obey the so-called minimal cooling scenario \citep{2004ApJS..155..623P}. In the Y+K EOS, degrees of hyperons and $K^-$ condensation are included, and the $\Lambda$-induced DU process occurs with $M_{\rm NS}=1.5~M_\odot$, but this effect is minor due to the presence of nucleon DU process. Note that $K^-$ condensation does not appear in the mass range we consider, regardless of the strength of interaction between nucleons and kaons~\citep{2026arXiv260509723B}. Although we select two EOS and limited mass ranges, their impacts on cooling curves are notably small in slow cooling scenario, as in a comprehensive study \citep{2017IJMPE..2650015L}.

In young neutron stars such as NS 1987A, the neutron ${}^1{\rm S}_0$ superfluidity in the inner crust is one of the important ingredients for its surface temperature. In this work, we consider six kinds of superfluidity: CCDK~\citep{1993NuPhA.555...59C}, SFB~\citep{2003NuPhA.713..191S}, WAP~\citep{1993NuPhA.555..128W}, GC~\citep{2008PhRvC..77c2801G}, GIPSF~\citep{2008PhRvL.101m2501G}, and T~\citep{1984PThPh..71.1432T} roughly in order of strength (the same as the right panel of Figure 6 in \cite{2020ApJ...898..125P}), in addition to no superfluidity (we refer to "All Unpaired"). Regarding the superfluidity in the core, we fix the "0.5a" model for ${}^3{\rm P}_2$ neutrons~\citep{2011PhRvL.106h1101P}, which reproduces the observed cooling rate of Cassiopeia A, while we fix the CCDK model for ${}^1{\rm S}_0$ protons~\citep{1993NuPhA.555...59C}. The corresponding maximum critical temperatures are $T_{c,n3}=5\times10^8~{\rm K}$ and $T_{c,p1}=6\times10^9~{\rm K}$, respectively. With the ages up to 200 yr, we confirm that the thermal luminosity remains largely unchanged for $T_{c,n3}\lesssim7\times10^8~{\rm K}$, since such a low critical temperature delays the onset of the pair breaking and formation process until after 200 yr (e.g., \cite{2009ApJ...707.1131P,2011PhRvL.106h1101P}). On the impact of proton superconductivity on young NSs, see Appendix \ref{app:a}.

In terms of envelope mass, we consider $\eta\equiv g_{14}\frac{M_{\rm env}}{M_{\rm NS}}=10^{-8}$ and $10^{-7}$, which connects the surface temperature and outermost crust temperature at $\rho=10^{10}~{\rm g~cm^{-3}}$, in accordance with the accreted envelope model of \cite{1997A&A...323..415P}. For $\eta\lesssim10^{-9}$, thermal luminosity is too low to account for the ALMA observations, even considering the uncertainties on extra heating mechanisms such as $^{44}{\rm Ti}$ radioactive heating~\citep{2019ApJ...886...51C,2020ApJ...898..125P}.

%Since the  ALMA observations require high envelope mass as $\eta\equiv g_{14}\frac{M_{\rm env}}{M_{\rm NS}}\gtrsim10^{-8}$~\citep{2020ApJ...898..125P,2023ApJ...949...97D}, the 

%Our interest of NS1987A's age of  $50~{\rm yr}\lesssim t\lesssim200~{\rm yr}$ could be involved in all kinds of nucleon superfluidity, since the heat from both inner-crust and core contributes to the surface luminosity (e.g., \cite{2009ApJ...707.1131P}). Thus, va
%Around the age of Various superfluidity 

%The ALMA observations require high envelope mass as $\eta\equiv g_{14}\frac{M_{\rm env}}{M_{\rm NS}}\gtrsim10^{-8}$

%(but see also JWST observation reported by \cite{2024Sci...383..898F}).

%%Observations and extrapolation
All sensitivity curves below retain the instrument response and remnant-background treatment of Paper I; they are a fixed \textit{Lynx}-like benchmark, not sensitivity forecasts for other missions. The subscript ``Lynx'' in Eq.~(\ref{eq:free}) identifies this benchmark. With the use of \textit{NSCool} code~\citep{2016ascl}, we systematically calculate thermal luminosities and compare with ALMA observations and sensitivity limits of a \textit{Lynx}-like observatory, the latter of which corresponds to the critical luminosity between non-detection and detection of NS 1987A. We take their concrete value from our previous work at $t=50~{\rm yr}$, $L(50)$ (see Table 1--6 in Paper I), assuming 
the conservative parameter sets of exposure time of 0.2 Ms and $3\sigma$ significance. The NS kick velocity is chosen to be $v_{\rm kick}=$500~${\rm km~s^{-1}}$ inferred from recent observations~\citep{2025ApJ...991..130L,2025A&A...704A.272T}. For the Y+K models, the correction of gravitational redshift is applied to the original values with the APR. We note that, in all cases, \textit{Lynx}-like sensitivity limits with $v_{\rm kick}\lesssim$500~${\rm km~s^{-1}}$ are higher than luminosities from cooling models at $t=50~{\rm yr}$.
%, as shown in Paper I. 

To consider the detection beyond 50 years, we extrapolate the \textit{Lynx}-like sensitivity limit as follows: Paper I verified the following time evolution of \textit{Lynx}-like benchmark sensitivity limits (see their footnote 23):
% Assuming that SNR 1987A freely expands with the original density distribution for $t\lesssim200~{\rm yr}$, the ejecta density becomes sparse as $t^{-3}$. In this case, 
%This uniform-density free expansion (UDFE) assumption is well justified by Paper I which showed that the ratio of ejecta density matches that of \textit{Lynx}'s sensitivity limits between $t=50$ yr and $t=56$ yr (see their footnote 23), which enables us to assume
%since the ejecta density becomes sparse as $t^{-3}$
\begin{eqnarray}
    L_{Lynx}(t)=L_{Lynx}(50)t^{-3}~\mathrm{for}~t\lesssim200~{\rm yr},~\label{eq:free}
\end{eqnarray}
which we assume for simplicity. This nearly mimics that the SNR 1987A freely expands with the original density distribution for $t\lesssim200~{\rm yr}$. To confirm the validity of such an approximation more, we calculate the \textit{Lynx}-like benchmark sensitivity limits at $t=60~{\rm yr}$ in the self-consistent manner same as in Paper I. Namely, they are obtained by synthesizing the 2047 \textit{Lynx}-like spectra of the faint inner region of SN 1987A, taking into account for the absorption pattern of cold ejecta based on the latest state-of-art magnetohydrodynamic
(MHD) simulation models broadly comparable with the observed morphology~\citep{oon20} \footnote[1]{
SN 1987A models published in \cite{oon20} have been extended up to 2047 ($t=60~{\rm yr}$) for this work, although they were originally calculated up to 2037.
%}
}.
%\eg{[EG: Please notice that for consistency with the estimates made in Paper I, I used the simulation by Salvatore of 2020, the same adopted for Paper I, to estimate the lynx limits a t = 60 yr.]}
Table~\ref{tab:lynx} presents a comparison of sensitivity limits with and without the approximation of Eq. (\ref{eq:free}) in the APR EOS. We found that \textit{Lynx}-like sensitivity limits are slightly overestimated with our approximation, leading to the longer putative detection age of NS 1987A. Thus, our above assumption is a \textit{conservative} one at least up to $t\lesssim60~{\rm yr}$, since the detection obtained in our analysis can be regarded as the upper limit value. This compensates for the high uncertainty on the structure of the ambient medium, currently unknown, in which the shock will expand in the next decades.

The detection age also depends on how thermal luminosity of NS 1987A evolves, that is, $L(t)\propto t^{-k}$ with power-law index $k$. A simple cooling theory shows $k\sim1/3-1/2$~\citep{2004ApJS..155..623P}, which is much smaller than the expansion rate of SNR as in Eq. (\ref{eq:free}). Thus, in the absence of rapid cooling, the declining absorption can bring thermal emission above the adopted sensitivity limit. To retain a concrete observing window, we use the same illustrative interval, 2045--2075 ($58\leq t\leq88~{\rm yr}$), corresponding to 30 years of operation. The plotted detection ages and detection/non-detection classifications are conditional on this sensitivity and window; a later start could miss a short-lived detectable phase. Labels referring to \textit{Lynx} in the existing figures denote this benchmark.

%These dates are assumptions of the benchmark calculation, not a launch schedule or a guaranteed mission lifetime.

\begin{table}[t]
    \centering
     \caption{\textit{Lynx}-like benchmark sensitivity limits in units of $10^{35}~{\rm erg~s^{-1}}$, at 2037($t=50~{\rm yr}$) and 2047($t=60~{\rm yr}$) calculated without and with the approximation of Eq. (\ref{eq:free}) (see texts for details).}
     \label{tab:lynx}
    \begin{tabular}{c|c|c|c}
    \hline\hline
NS mass  &     2037  & 2047 & 2047 (w/ approx.)\\
  \hline
  $1.20~M_\odot$       & 1.38 & 0.557 & 0.799 \\
  $1.40~M_\odot$       & 1.42 & 0.573 & 0.822 \\
  $1.60~M_\odot$       &  1.45 & 0.580 & 0.839 \\
    \end{tabular}
     \label{tab:L35}
\end{table}

%\textcolor{cyan}{

%}
%Assuming an isothermal non-superfluid crust and isothermal non-superfluid core with the temperature $T_{\rm crust}$ and $T_{\rm core}$, respectively, it is given as 
% \begin{eqnarray}
%     k = 
%     \begin{cases}
%         1/2 & T_{\rm crust}>T_{\rm core} \\
%         1/3 & T_{\rm crust}=T_{\rm core}~\mathrm{and~w/o~ DU}\\
%         1/2 & T_{\rm crust}=T_{\rm core}~\mathrm{and~w/~DU}
%     \end{cases}~,\label{eq:k}
% \end{eqnarray}

%\ka{もう少しpower-lawの議論を書かないといけないかな。Appendix Aも同様。}

%If the superfluidity is considered, these $k$ value significantly increases. In the case of $T_{\rm crust}=T_{\rm core}$, this is due to the enhanced cooling of PBF by ${}^3{\rm P}_2$ neutron superfluidity~\citep{2011PhRvL.106h1101P}. For the case of $T_{\rm crust}>T_{\rm core}$, this is mainly due to the significant reduction of neutron specific heat due to pairing of ${}^1{\rm S}_0$ neutron superfluidity, but not the PBF by~\citep{2009ApJ...707.1131P}. These are discussed in the Appendix.

%Except for NS 1987A and magnetars, most observed isolated NSs are thought to be isothermal because of their old age (>300 yrs) enough to complete the heat diffusion, implying $T_{\rm crust}=T_{\rm core}$ \cite{2011MNRAS.411.1977Y}. However, young NSs as a representative of NS 1987A should have a hot crust due to its youth, that is, $T_{\rm crust}>T_{\rm core}$. 

%assuming isothermal approximation~(e.g., \cite{2004ApJS..155..623P,2011PhRvL.106h1101P}):
%\begin{eqnarray}
%    L(t)=t^{-\alpha},~\label{eq:l}
%\end{eqnarray}
%\ka{ここはまだ途中。。。}

\section{Results}
\label{sec:res}

\subsection{Cooling curve with young NSs}

% \begin{figure}[t]
%     \centering
%     %\includegraphics[width=\linewidth]{./}
%     \caption{Various cooling curves and effective Lynx sensitivity limits by \textit{Lynx}. \\ \adcom{\ka{泉くん、cooling curveとLynx curveを比べた図を載せて。モデルは多く載せれるといいが、その辺は要相談。}}
%     }
%     \label{fig:mass}
% \end{figure}

\begin{figure}[t]
    \centering
    \includegraphics[width=1.0\linewidth]{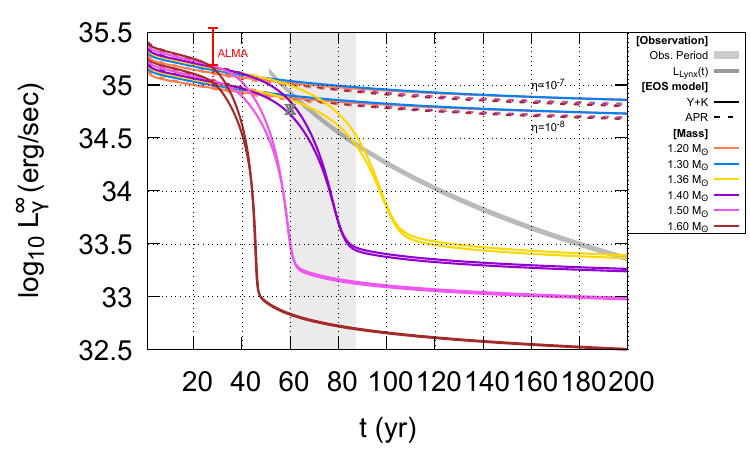}
    \caption{Mass dependence of cooling curves and \textit{Lynx}-like benchmark sensitivity limits with Y+K and APR EOS. Sensitivity limits without the approximation are also plotted as triangle symbols, taking into account the uncertainties of EOS and mass.}
    %\adcom{\ka{Lynx Windowの説明（30年設定の理由）}}
    \label{fig:mass}
\end{figure}

Fig.~\ref{fig:mass} presents the EOS and mass dependence of the comparison. In APR EOS, all NSs cool slowly according to the minimal cooling scenario, leading to the detection of NS 1987A by a \textit{Lynx}-like observatory around $t\approx58$ yrs with $\eta=10^{-7}$ while $t\approx63$ yrs with $\eta=10^{-8}$. In Y+K EOS, NSs cool faster than those with the APR EOS, except with 1.20 \& 1.30$M_\odot$ stars, due to the presence of the DU processes. We found that most of the rapid cooling models do not intersect with the sensitivity curves by a \textit{Lynx}-like observatory, implying the non-detection of NS 1987A by a \textit{Lynx}-like observatory. Thus, the DU process generally disfavors detection within the adopted benchmark window.

The exceptional models are those with $1.36~M_\odot$ stars, where cooling curves are temporarily located in higher-luminosity regions than the sensitivity curves by a \textit{Lynx}-like observatory. In this case, NS 1987A is detectable for some years, like a \textit{shooting star}\footnote[2]{
Although this is based on an approximation of Eq. (\ref{eq:free}), even without imposing it, \textit{shooting star} scenario is realized in Y+K EOS with a slight increase of NS mass.
%although the corresponding NS mass must slightly increase.
}. This is because the mass is close to the threshold mass $M_{\rm DU}$, implying that the volume where the DU process is active is not large enough to significantly affect surface luminosity (e.g., \cite{2004ARA&A..42..169Y}). Although how close they are strongly depends on EOS~\citep{2017IJMPE..2650015L,2020A&A...642A..42S}, such a "weak" DU process may realize the detection of NS 1987A.

\begin{figure}[t]
    \centering
    \includegraphics[width=1.0\linewidth]{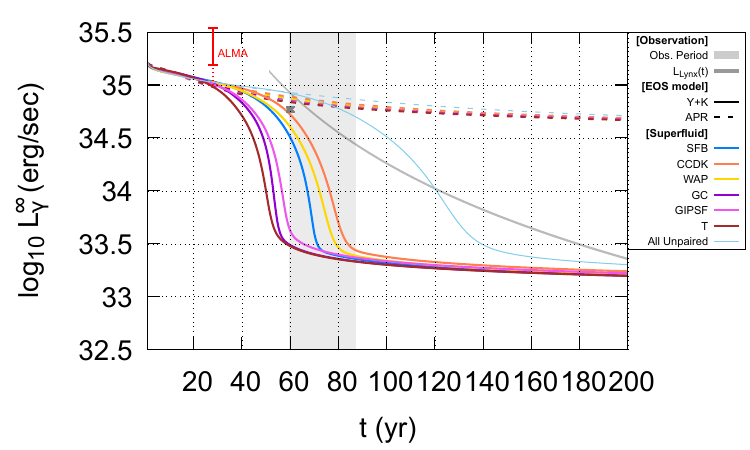}
    \caption{Same as Fig. \ref{fig:mass}, but neutron $^{1}S_0$ superfluid-model dependence of cooling curves assuming $1.40~M_\odot$ stars.
    }
    \label{fig:crustsf}
\end{figure}

Fig.~\ref{fig:crustsf} presents the impact of crust superfluidity on the detection of NS 1987A, assuming $1.40~M_\odot$ stars. As already reported by \cite{2020ApJ...898..125P}, the crust superfluidity reduces thermal luminosity of young NSs because of the suppression of the neutron specific heat~\citep{2009ApJ...707.1131P}. This becomes more significant with the Y+K compared to the APR, which allows the DU process to occur; one can see that stronger crust superfluidity leads to earlier rapid cooling. Except for the case without crust superfluidity, one can see no intersection between the cooling curve and the sensitivity curve by a \textit{Lynx}-like observatory, showing the non-detection of NS 1987A with the DU process. 

\begin{figure*}[t]
    \centering
    \begin{minipage}{0.49\linewidth}
    \includegraphics[width=\linewidth]{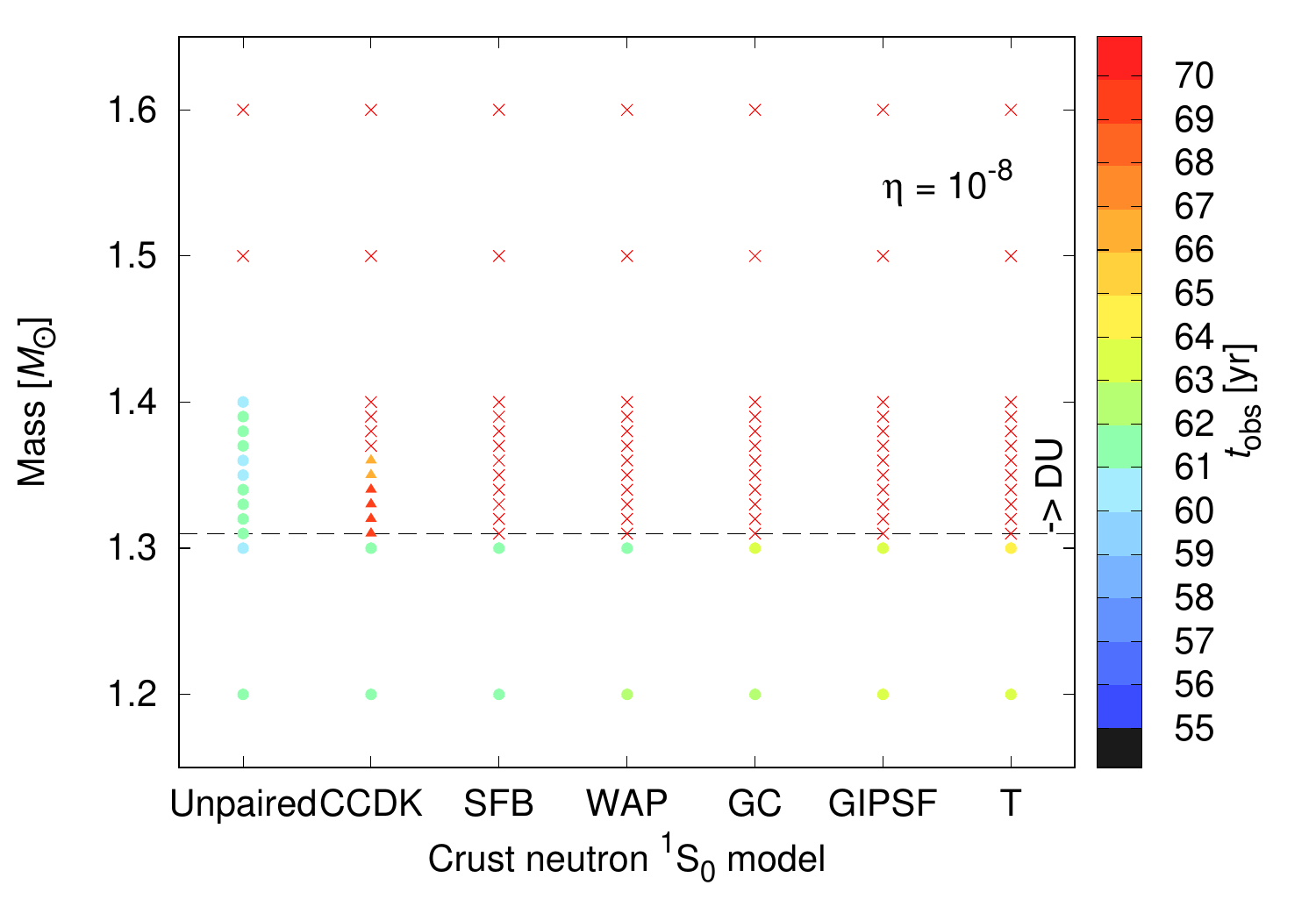}
    \end{minipage}
     \begin{minipage}{0.49\linewidth}
    \includegraphics[width=\linewidth]{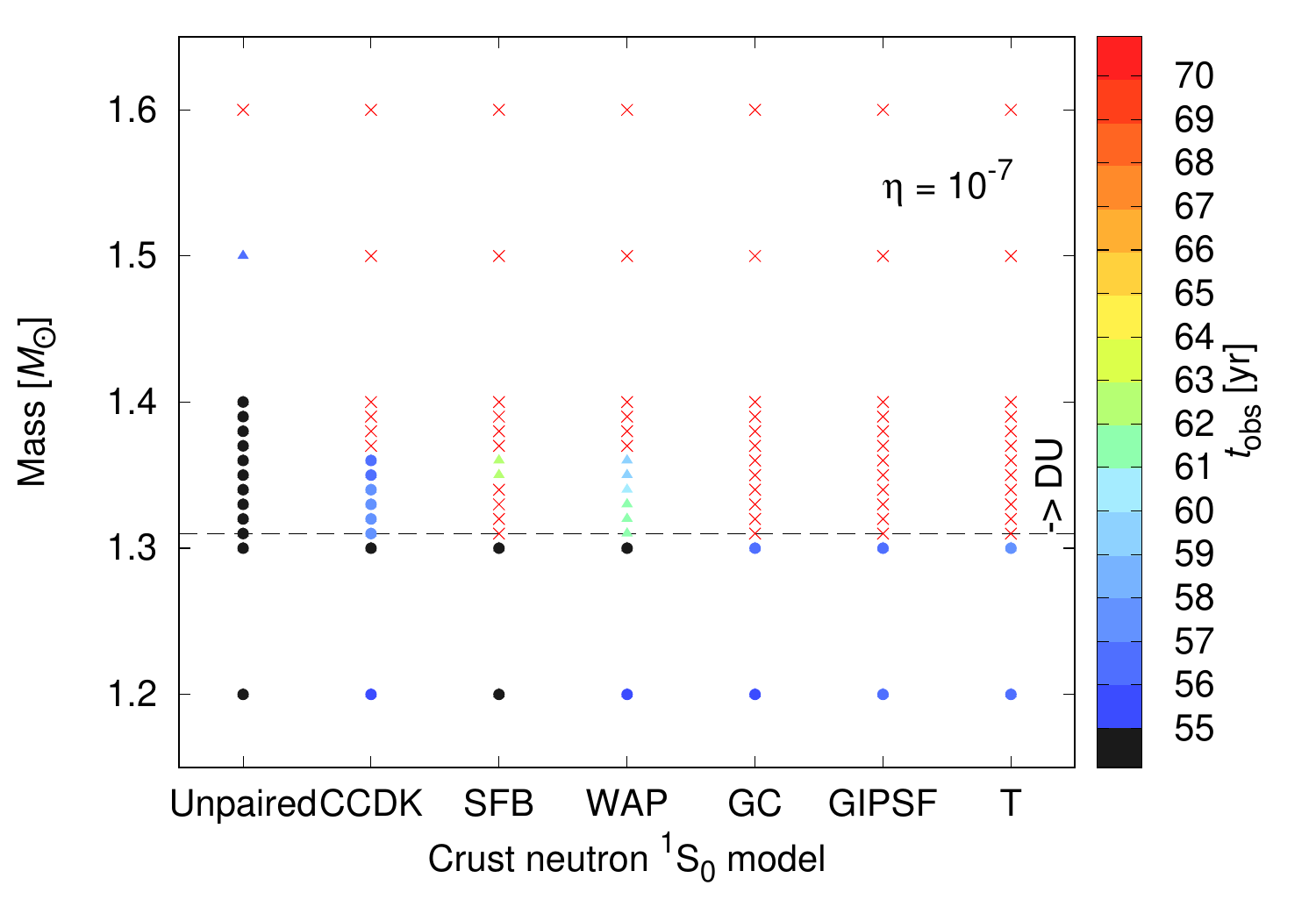}
    \end{minipage}
    \caption{NS mass, $\eta$, and crust superfluid model dependences of the detection or non-detection of NS 1987A with the Y+K EOS: $\bullet$ denotes the detection while $\times$ denotes non-detection within 88 years after the birth. $\triangle$ denotes the shooting-star scenario. The contour indicates the age when NS 1987A is detected by a \textit{Lynx}-like observatory. The dashed line corresponds to the critical mass of the DU process with the Y+K EOS, $M_{\rm DU}=1.31~M_\odot$.}
\label{fig:summary}
\end{figure*}

 While the APR EOS realizes the detection in all parameter regions we considered, the Y+K EOS does not always do so due to the presence of the DU process. Fig. \ref{fig:summary} displays how the detectability of the NS 1987A changes with the Y+K EOS. If the NS mass is lower than the critical mass of the DU process, all models show the detection, as with the APR, but for $M_{\rm NS}\gtrsim M_{\rm DU}$, the detectability depends on the crustal superfluidity. In general, a stronger crust superfluid model tends to make the DU cooling more efficient due to the reduction of neutron specific heat, finally shortening the thermal relaxation timescale~\citep{2001MNRAS.324..725G}. One may think that $t_{\rm obs}$ is smaller with $M_{\rm NS}/M_\odot$=1.35 and 1.36 compared with that with $M_{\rm NS}/M_\odot$=1.31--1.34 with weak superfluid models (CCDK, SFB, and WAP). This seems to be attributed to the difference in crust thickness, which is connected to the thermal relaxation timescale. In fact, the crust thickness is longer for $1.35~M_\odot$ stars by 1.1\% than that for $1.34~M_\odot$ stars.
 
 If $M_{\rm NS}-M_{\rm DU}$ becomes larger, due to the large-volume core of DU cooling, it becomes more impossible to detect NS 1987A due to rapid cooling before the putative detectable age (see also Fig. \ref{fig:mass}). The higher envelope mass also makes detection more possible, but except for mass close to $M_{\rm DU}$, it has little impact on the criterion of detection and non-detection unless we consider a small envelope mass enough not to account for the ALMA observations.
 
\subsection{Detection age of NS 1987A v.s. Relaxation time}

As we see the above, by comparing cooling curves with sensitivity limits by a \textit{Lynx}-like observatory that consider the X-ray absorption due to supernova ejecta, one can know the detection age of NS 1987A, $t_{\rm obs}$. Also, we calculate the effective thermal relaxation timescale $t_{\rm relax}$, defined as the age when the cooling rate $|\frac{d \ln L^\infty_\gamma}{d \ln t}|$ is maximum for $1<t<200~{\rm yr}$, which mimics when isolated NSs become isothermal \citep{1994ApJ...425..802L}. Most isolated NSs observed are thought to be isothermal, i.e., $t_{\rm obs}>t_{\rm relax}$, such as the case of Cassiopeia A~(e.g., \cite{2011MNRAS.411.1977Y}), but except for our target young neutron star, NS 1987A: If $t_{\rm obs}<t_{\rm relax}$, the inner crust should have hotter than the core, implying that the neutrino cooling in the inner crust plays the most important role on the surface luminosity. In other words, NS 1987A is a potentially unique target for probing the properties of inner-crust matter using future luminosity observations. 

%All Unpaired models show higher luminosity compared with cooling models with crust superfluidity, implying the possibility of detection. 

\begin{figure}[t]
    \centering
    \includegraphics[width=1.0\linewidth]{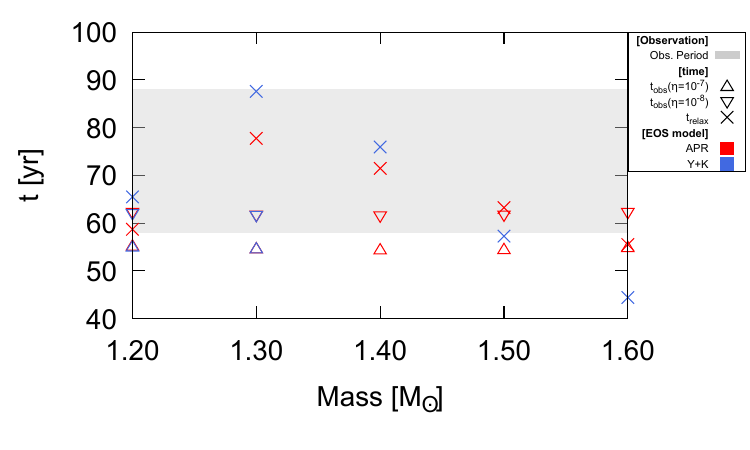}
    \caption{Mass dependence of observation time and relaxation time with APR and Y+K EOSs, assuming neutron $^1{\rm S}_0$ superfluid model as the CCDK. The observational ages with Y+K EOS for $1.40, 1.50$ and $1.60M_{\odot}$ are $t_{\rm obs}\approx202~{\rm yr}$ regardless of $\eta$ values, respectively.
    %201.6yr, 202.0yr, 201.8yr (for $\eta = 10^{-7}$) and 202.3yr, 202.4yr, 201.9yr (for $\eta = 10^{-8}$), respectively.\\
}
    \label{fig:tobs1}
\end{figure}

Fig. \ref{fig:tobs1} presents EOS and mass dependence of $t_{\rm obs}$ and $t_{\rm relax}$ with the CCDK of crust superfluidity, which implies $t_{\rm obs}\sim60~{\rm yr}$ except for Y+K EOS with $M_{\rm NS}\gtrsim1.40~M_\odot$, where the DU process operates.
The behavior of $t_{\rm relax}$ is complicated by superfluidity \citep{2001MNRAS.324..725G}, but, except for 1.20$M_\odot$ stars, higher-mass models tend to yield smaller $t_{\rm relax}$ (this feature can also be seen with APR). This derives from slightly early cooling around 100 yr even with the slow cooling scenario (e.g., \cite{2004ARA&A..42..169Y,2017IJMPE..2650015L,2019PTEP.2019k3E01D}). 

One can also see two cases of $t_{\rm obs}<t_{\rm relax}$ and $t_{\rm obs}>t_{\rm relax}$, which imply that NS 1987A is not isothermal (where the crust is hotter than the core) and isothermal at the detection age, respectively. For instance, if the DU process occurs (With Y+K EOS and $M_{\rm NS}\gtrsim1.40~M_\odot$), it corresponds to the latter case. On the other hand, the APR models with $M_{\rm NS}=1.30-1.50~M_\odot$ show the former case. Therefore, future luminosity observations of NS 1987A by a \textit{Lynx}-like observatory can probe the physical properties in either the inner crust or the core, respectively, depending on model parameters.

%all models result in $t_{\rm obs}<t_{\rm relax}$For instance, focusing on the APR EOS,  for $M_{\rm NS}\lesssim1.4~M_\odot$ while $t_{\rm obs}<t_{\rm relax}$ for $M_{\rm NS}\gtrsim1.5~M_\odot$, 

\begin{figure}[t]
    \centering    \includegraphics[width=1.0\linewidth]{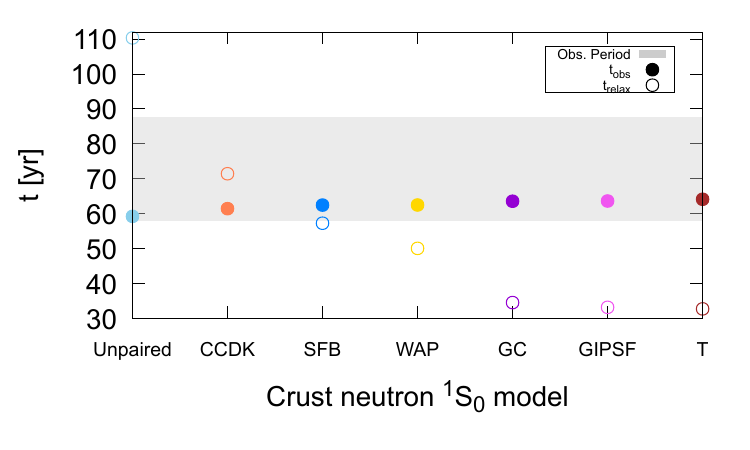}
    \caption{Neutron $^{1}S_0$ Superfluid dependence of observation time and relaxation time with APR EOS at $1.4M_{\odot}$.}
    \label{fig:tobs2}
\end{figure}

The most important ingredient on $t_{\rm relax}$ under the minimal cooling scenario is the crust superfluidity, as shown in Fig. \ref{fig:tobs2}: The stronger the crust superfluidity is, the shorter $t_{\rm relax}$ is. Meanwhile, $t_{\rm obs}$ does not drastically increase by crust superfluid models. Therefore, if the crust superfluidity is weak, such as the CCDK model,  one can show $t_{\rm relax}>t_{\rm obs}$, implying that NS 1987A is a potential site to probe crustal properties.

%\subsection{What does the observed cooling slope tell us?}

%\ka{泉くんともう少し要相談。Appendix と関係。}
%%実際の超流動モデルとの乖離

%%

%\subsection{The cooling curves and implications for superfluidity}

\section{Discussion and Conclusion}
\label{sec:con}

We investigated how the thermal evolution of the putative NS 1987A connects future luminosity measurements to its interior physics. For the adopted\textit{Lynx}-like sensitivity and observing window, rapid neutrino cooling, such as the DU process, generally disfavors detection because it reduces the surface luminosity before the thermal emission becomes accessible. However, for a mass close to the threshold for rapid cooling, the star can remain detectable for a short interval, as in the shooting-star scenario. These are conditional observational diagnostics, rather than predictions for the launch of a specific telescope.

In the standard cooling scenario, NS 1987A can be detected at $t\sim60~{\rm yr}$ if a facility reaching the adopted\textit{Lynx}-like sensitivity is operating at that epoch. This is because (i) the cooling rate is smaller than the decrease of \textit{Lynx}-like sensitivity limit, assuming that SNR 1987A is freely expanding, and (ii) ALMA observations by \cite{2019ApJ...886...51C} imply high-mass envelope with $\eta\gtrsim10^{-8}$ at $t\simeq28~{\rm yr}$, resulting in high thermal luminosity enough to enable the future detection of NS 1987A.

Table~\ref{tab:missions} compares representative X-ray facilities with the adopted benchmark. Angular resolution is essential as well as effective area: the bright equatorial ring is only about 1.6 arcsec across~\citep{oon20}, so subarcsecond imaging is needed to reduce its contamination in an extraction region around the compact object. This is not a universal sharp resolution threshold, since the point-spread-function wings, source position, and remnant brightness also matter. Chandra provides such imaging but has a much smaller effective area. AXIS would have increased the count rate, but its broader point-spread function would have mixed more remnant emission into the central region; moreover, it became ineligible for the Probe selection in March 2026~\citep{AAAC_Report2026}, and NASA subsequently selected the far-infrared mission PRIMA~\citep{NASA_PRIMA2026}. NewAthena offers a large effective area, but its $\sim9$ arcsec imaging cannot spatially separate the central source from the ring. Nevertheless, among the available satellites in the 2040s, NewAthena would be the most promising candidate to detect thermal emission from NS 1987A. Its X-IFU spectroscopy could constrain the shocked plasma and help search for an additional central component in the integrated spectrum, although isolating thermal NS emission would remain model dependent. HUBS, designed for diffuse soft X-ray spectroscopy, likewise cannot spatially isolate the central source. Thus, these alternatives offer complementary constraints but do not provide a demonstrated replacement for the benchmark thermal detection. In particular, evidence for a nonthermal PWN component, as in Chandra--NuSTAR studies~\citep{2021ApJ...908L..45G,gmo22}, is distinct from the surface thermal luminosity needed for our cooling diagnostic.

\begin{table*}[t]
\centering
\caption{Representative X-ray capabilities and mission status (October 2026).}
\label{tab:missions}
{\small
\begin{tabular}{llll}
\hline\hline
Facility & Angular resolution & Area at 1 keV ($ {\rm cm}^{2}$) & Launch / status \\
\hline
Chandra/ACIS-S & $<0.5''$ (mirror FWHM) & $\sim555$ (nominal, BI) & 1999; operating \\
\textit{Lynx}-like benchmark & $0.5''$ (HPD) & $\sim20{,}000$ (mirror) & No scheduled launch; unfunded concept \\
AXIS concept & $1.5''$ (on-axis HPD) & $\sim3{,}900$ (incl. detector) & Ineligible in 2026 \\
NewAthena & $<9''$ (on-axis HEW) & $>11{,}000$ (mirror req.) & 2039 target; adoption pending \\
HUBS concept & $<60''$ (HPD) & $\sim500$ (FoV-averaged) & Proposed; launch not confirmed \\
\hline
\end{tabular}
\par\vspace{2mm}
\begin{minipage}{0.96\textwidth}\footnotesize
Note. HPD and HEW denote the diameter enclosing half the energy; FWHM is a different measure. Areas are published reference values, not a uniform set of detector-folded, epoch-specific responses. The Chandra value is the nominal ACIS-S back-illuminated (BI) value, not its 2026 area; contamination reduces its soft response with time. \textit{Lynx} and NewAthena values refer to mirror areas, before detector/filter losses. The AXIS value is the updated 2025 on-axis baseline, superseding the $4{,}200~{\rm cm}^{2}$ value on the mission website. NewAthena entries are optical design requirements (req.); ideal-optics ray tracing predicts $\sim1.24~{\rm m}^{2}$ at 1 keV, about 62\% of the \textit{Lynx} mirror benchmark, rather than a detector-inclusive area or a sensitivity ratio. Future dates are targets, not guaranteed launches. The current English HUBS roadmap lists pathfinder and engineering development without a launch date. Sources: Chandra X-ray Center~\citep{CXC_PocketGuide}; NASA's \textit{Lynx} concept reassessment~\citep{2026arXiv260900033L}; the AXIS camera team~\citep{AXIS_Camera2025}, the AAAC status report~\citep{AAAC_Report2026}, and NASA's Probe selection~\citep{NASA_PRIMA2026}; NewAthena optics studies~\citep{NewAthena_Optics2024,INAF_BEaTriX} and ESA's current launch target~\citep{ESA_NewAthena}; HUBS mission description~\citep{HUBS_Mission}.
\end{minipage}}
\end{table*}

Our cooling curves therefore remain useful even if \textit{Lynx} is never launched: a measured thermal luminosity and age, compared with the relaxation timescale, can distinguish an interior that has thermally relaxed from a hotter crust and test crustal superfluidity within the model uncertainties discussed here. Whether a particular facility can supply that luminosity must be assessed with its full energy-dependent response, exposure, ejecta absorption, and remnant contamination. A nondetection is informative only when its thermal-luminosity upper limit reaches the relevant cooling models. We do not transfer the numerical detection ages to NewAthena, AXIS, or HUBS by scaling their areas; such forecasts would require new spectral simulations. The benchmark therefore provides a reference for assessing future thermal measurements against the cooling models.

We also discussed the implications when NS 1987A is detected. It is found that NS 1987A is likely to be isothermal at the detection, unless the neutron superfluidity in the crust is weak. This implies that the observed luminosity and age of NS 1987A can extract the information on the NS core, but not the inner crust, except for the CCDK model of $^1{\rm S}_0$ neutron superfluidity. 

 The $^1{\rm S}_0$ neutron superfluidity has been discussed in terms of condensed-matter physics (for reviews, see \cite{2019EPJA...55..167S,2020PrPNP.11103739O}). Under the standard mean-field (BCS) theory, $T_c$ value is around 20 GK~\citep{2019NatSR...918477T}, which is also supported by the lattice Monte-Carlo calculation~\cite{2009PhRvC..79e4002A}. This implies that NS 1987A could not be a probe for the properties inside the inner crust. However, if the beyond mean-field effects, such as the pairing fluctuation and many-body effects, are considered, the efficiency of cooling inside the inner crust will be changed, such as the enhancement of specific heat~\citep{2020PhRvC.102f5802I}, and reduction of $T_c$ due to the so-called Gork'ov and Melik-Barkhudarov (GMB) effects~\citep{GMB}. To be slightly optimistic, investigating more sophisticated neutron superfluid models in the inner crust beyond BCS theory may make the detection of NS 1987A lead to a unique probe of crustal matter. 

%\textcolor{cyan}{
The present work serves as a general approach to specify the physics of NS 1987A from future detection of thermal emission. Although the physical link between cooling, relaxation, and luminosity does not depend on the observing facility, the detection ages are conditional on our benchmark; it should be mentioned about the following two issues: One is the limitation of the number of cooling models, in particular with the fast cooling scenario. Then, it would be worth doing more systematic calculations to judge the detectability, such as Fig. \ref{fig:summary} with different EOSs (threshold mass for the onset of the DU process) and stellar mass. Another issue is to assume Eq. (\ref{eq:free}) for $t>50~{\rm yr}$, whose approximation becomes worse as NS 1987A gets older due to an increase in the swept mass from the circumstellar medium. Our imposed approximation 
is somewhat conservative, as shown in Table~\ref{tab:lynx}. Fortunately, this means that the detection would be achieved more easily. Access to sensitive X-ray observations during the thermal-relaxation epoch would therefore be particularly valuable, irrespective of the mission that provides them.
%at least for the definitive detection age $t=60~{\rm yr}$  which means that the detection must meet with $\lesssim$
%seems to hold 
%Moreover, while the ejecta in the core of SNR 1987A is almost flat, that in the envelope has a steep density structure as $\rho_{ej}\propto t^{-3}(r/t)^{-n}$ with the power-law index $n\sim$10--12~\citep{1982ApJ...258..790C,1999ApJS..120..299T}. This may be attributed to the overestimation of \textit{Lynx} sensitivity limits shown in Table 1.

To determine the detection age of NS 1987A more accurately, systematic calculations of sensitivity limits with realistic ejecta profiles are necessary in the future. However, they take a huge computational cost due to time-to-time spectrum synthesis with 3D MHD models \citep{2025A&A...699A.305O}. Moreover,  the morphology described by 3D MHD models would still be affected by the large uncertainty on the structure and composition of the outer ambient medium.

%and thus, the exact determination of the detection age of NS 1987A is left for future study

%necessary for a more accurate prediction of the detection age of NS 1987A. Moreover, the absorption effects of the ejecta are still affected by the uncertainty on the structure and composition of the outer ambient medium, which will be 

%hich would still be affected by the uncertainty on the structure and composition of the outer ambient medium, currently unknown. The more accurate prediction of the detection age of NS 1987A is left for future study.
%}

%We plan to calculate the sensitivity limit of high-resolution X-ray satellites and refine the current results. As the higher the detector's effective area and allocated exposure time they are, it is expected to be an earlier detection of NS 1987A, which would justify the free-expansion approximation.

Let us briefly mention the putative impact of the internal magnetic field which we did not consider. If the dipole magnetic field at the pole is as high as $\sim3\times10^{15}~{\rm G}$, the  Joule heating increases thermal luminosity of NS 1987A to $L^\infty_\gamma\sim5\times10^{35}~{\rm erg~s^{-1}}$~\citep{2013MNRAS.434..123V}, making the detection of NS 1987A earlier. Because the free precession period that slightly differs from the rotational one can be measured in magnetars enclosed in SNR~\citep{2026PASJ..tmp...21M}, one can judge whether NS 1987A is a magnetar or not before future deep thermal X-ray observations\footnote[3]{Note that the pulse associated with the free precession appears as the hard X-ray with $\sim10~{\rm keV}$, which would not be so sensitive to absorption effects by SNR ejecta.}. If NS 1987A turns out to be a magnetar, our scenario to detect thermal emission from NS 1987A would become more testable.

%

%In this work, we did not consider the magnetic field The detectable age of $\sim60~{\rm yr}$ It should be also mentioned that 

%\ka{他の観測器の話（長瀧さんと要相談）　例えば、AXISではどうか？が必ず入りそう。大雑把にeffective areaでみるしかない？（１−２文？or 書かない？）}

%and its implications on the physics of neutron stars, such as the EOS, mass, and superfluidity. 

%%

\section*{Acknowledgements}

We thank T. Muto for providing his EOS data. We also thank D. Patnaude and P. Slane for helpful discussion on \textit{Lynx}, and R. Hirai on SNR ejecta. This work is supported by JSPS KAKENHI (JP23K25874, JP25H00675, JP25K17403, and JP26K17149). S.N. is also supported by the JST ASPIRE Program ”RIKEN-Berkeley mathematical quantum science initiative”. 

%A.I. is supported by JSPS KAKENHI Grant Number JP22K14034 and JP26K17149.S.N. is supported by JSPS KAKENHI Grant Number JP23K25874 and JP25H00675.
{\it Software:} Gnuplot~\citep{2022gnuplot}
\bibliography{ref}{}
\bibliographystyle{aasjournal}

\appendix

\section{On the impact of proton superconductivity on NS 87A}
\label{app:a}

\begin{figure}[h]
    \centering
    \includegraphics[width=0.6\linewidth]{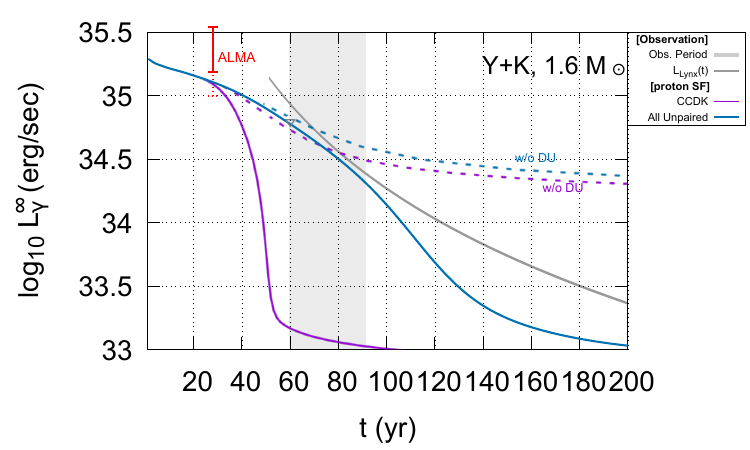}
    \caption{Cooling curves with and without CCDK proton superconductivity, assuming the CCDK neutron $^1{\rm S}_0$ superconductivity model, Y+K EOS and 1.6$M_\odot$ stars. Observational properties plotted are the same as Fig. 1, but assuming the adopted mass and EOS.}
    \label{fig:pSF}
\end{figure}

Around the age of NS 1987A, the proton superconductivity should have already been realized in the core. As time passes, thermal relaxation proceeds, and neutrino cooling in the core eventually becomes dominant. Here, we discuss the role of proton superconductivity on cooling curves of young NSs.

Fig. \ref{fig:pSF} shows the cooling curves with and without CCDK proton superconductivity, fixing the Y+K EOS with $1.6~M_\odot$. In the fast cooling case where the DU process occurs, the proton superconductivity makes the cooling timescale shorter by $\sim$50 yr. In the minimal cooling case (i.e., when the DU process is artificially switched off), the behavior of cooling curves is similar, but the detection age by \textit{Lynx}-like observatory is changed by $\sim10~{\rm yr}$. Considering that the CCDK is a relatively strong model~\citep{2015PhRvC..91a5806H,2021PhRvC.103b5807L}, this 10-yr difference may be a rough upper estimate of the uncertainty in the detection age associated with proton superconductivity, at least under a minimal cooling scenario.

The reason why proton superconductivity affects cooling curves of young NS is mainly due to the change of electron thermal conductivity in the core, which is given by $\kappa_e = \frac{\pi^2k_B^2Tn_e}{3m_e^*\nu_e}$, where $k_B$ is the Boltzmann constant, $n_e$ is the electron number density, and $m_e^*$ is the effective electron mass at the Fermi surface. $\nu_e$ denotes the effective collision frequency, and neglecting the contribution from muons, it is expressed as $\nu_e\simeq\nu_{ee}+\nu_{ep}$, where $\nu_{ee}$ and $\nu_{ep}$ denote the contributions from electron-electron and electron-proton scattering, respectively. In the presence of proton superconductivity, $\nu_{ep}$ is suppressed due to fewer protons available to scatter. $\nu_{ee}$ is also suppressed by so-called Meissner screening~\citep{2007PhRvD..75j3004S,2015PhRvC..91b5806B}. Both effects therefore increase $\kappa_e$, which shortens the cooling timescale as expected in Fig.~\ref{fig:pSF}. We note that how the cooling timescale is changed by proton superconductivity depends \textit{nontrivially} on NS mass, since it is proportional to the crust thickness squared and inversely proportional to the relativistic factor cubed~\citep{1994ApJ...425..802L,2001MNRAS.324..725G}. Generally, both the crust thickness and the relativistic factor decrease as the NS is heavier. Moreover, \cite{2020A&A...642A..42S} shows that, in the NS mass regions just above $M_{\rm DU}$, the thermal relation becomes much slower under non-superfluid cooling models.

\end{document}